\documentclass[oneside]{elsart}
\usepackage{pslatex}
\usepackage[T1]{fontenc}
\usepackage[latin1]{inputenc}
\usepackage{graphicx}
\usepackage[numbers]{natbib}


\begin{document}
\journal{JQSRT}

\begin{frontmatter}

\title{Fast Computation of Voigt Functions via Fourier Transforms}

\author{Marcus H. Mendenhall\thanksref{mfel}}

\thanks[mfel]{Supported by grant FA9550-04-1-0045 from the DOD MFEL Program}

\address{Vanderbilt University Free Electron Laser Center, P. O. Box 351816 Station B, Nashville, TN 37235-1816, USA} \ead{marcus.h.mendenhall@vanderbilt.edu}

\begin{abstract}

This work presents a method of computing Voigt functions and their derivatives,  to high accuracy, on a uniform grid.  It is based on an adaptation of Fourier-transform based convolution.  The relative error of the result decreases as the fourth power of the computational effort.  Because of its use of highly vectorizable operations for its core, it can be implemented very efficiently in scripting language environments which provide fast vector libraries.  The availability of the derivatives makes it  suitable as a function generator for non-linear fitting procedures.

\end{abstract}


\end{frontmatter}

\section*{Introduction}

The computation of Voigt line profiles is an issue which has been
dealt with over a long time in the literature \citep[e.g.]{drayson1976,pierluissi1977,karp1978,schreier1992}.
Nonetheless, it remains a computationally interesting problem because
it is still fairly expensive to compute profiles to high accuracy.
This paper presents a method which is fast and very simple to implement.
It is similar to the method of \citep{karp1978}, but capable of much
higher precision for a given computational effort. More importantly,
the method described here computes not only the Voigt function itself,
but its derivatives with respect to both the Gaussian and Lorentzian
widths, which are helpful for non-linear curve fitting.

Because the method is based on Fourier transforms, and generates a
grid of values for the function in a single operation, it is particularly
suitable for (but not restricted to) use in scripting environments,
where Fast Fourier Transforms (FFTs) and other vector operations are
available as part of the language library, and execute at very high
speed. Thus, in Matlab\textregistered, Python (with Numeric or numpy),
Mathematica\textregistered, Octave, and other similar environments,
it may find particular applicability. Such environments provide a
trade-off of speed for flexibility, and this method allows one to
take a much smaller speed penalty than would be exacted using other
algorithms implemented in the scripting language.

A method such as this one, which contains no inherent approximations,
has advantages over other approximate methods in that it can be adapted
very easily to the desired level of precision of the task at hand.
Selecting the density of the computational grid and the length into
the tails to which the computation is carried out sets the accuracy.
The grid density does not affect the accuracy of the values this produces
directly, but does affect the accuracy to which interpolation may
be carried out between the computed points. The distance into the
tails to which the profile is computed does affect the accuracy of
the step (defined below) which corrects for periodic boundary conditions,
and it converges as the fourth power of this distance.

As an aside on notation, most papers working with the Voigt function
historically have defined it in terms of a single parameter, the ratio
of the Lorentzian to Gaussian width. This work, presents the results
in a slightly different form, which is more useful for direct computation.
The equations below are computed in terms of the Lorentzian width
(which I call $\alpha$) and the standard deviation of the Gaussian
distribution $\sigma$. The parameter $y$ in Drayson \citep{drayson1976}
is $\alpha/(\sqrt{2}\sigma)$ in my notation. In this form, the transforms
produce functions fully scaled and ready for direct interpolation.

\section*{Theory}

The Voigt function is a convolution of a Lorentzian profile and a
Gaussian, \begin{equation}
V(\alpha,\sigma;\, x)\propto\int_{-\infty}^{\infty}\frac{e^{-t^{2}/2\sigma^{2}}}{(x-t)^{2}+\alpha^{2}}\, dt\label{eq:conv}\end{equation}
and can be easily written down in Fourier transform space using the
convolution theorem:\begin{equation}
\tilde{V}(\alpha,\sigma;\, k)=\exp\left[-\sigma^{2}k^{2}/2-\alpha\left|k\right|\right]\label{eq:convfft}\end{equation}

Also, of great importance to using this in fitting procedures, the
derivatives of this function with respect to its parameters can be
computed:\begin{equation}
\frac{\partial\tilde{V}}{\partial\alpha}=-\left|k\right|\tilde{V}\label{eq:ddalpha}\end{equation}
and\begin{equation}
\frac{\partial\tilde{V}}{\partial\sigma}=-\sigma k^{2}\tilde{V}\label{eq:ddsigma}\end{equation}

and since the differentiation in these cases commutes with the transform,
these are the transforms of the appropriate derivatives of the function
itself. 

Note that, since this transform method generates functions with a
fixed area, these are the derivatives with respect to the widths at
fixed area, rather than at fixed amplitude. This implies that fitting
carried out using functions computed this way is most appropriately
done using $\alpha$, $\sigma$, and area as parameters.

This result is exactly correct in the full Fourier transform space.
For practical computation, though, one wishes to reduce this into
something which is computed rapidly on a discrete lattice using Fast
Fourier Transform (FFT) techniques, and then interpolated between
the lattice points. The difference between the full continuous transform
and the discrete transform is, of course, that the function produced
by a discrete transform is periodic. In effect, by discretely sampling
the series in Fourier space, one is computing, instead of the exact
convolution, a closely related function which has had periodic boundary
conditions applied. This affects the shape of the tails of the distribution,
but in a way which is fairly easily fixed. 

First, when doing the discrete transforms, it is necessary to decide
how far out in $k$-space it is necessary to have data. In general,
one wants to assure the function is nicely band-limited, which means
no significant power exists at the highest $k$. Practically speaking,
setting the argument of the exponential in eq. \ref{eq:convfft} to
something like --25 at the boundary means the highest frequency component
is $e^{-25}$ or about $10^{-11}$ of the DC component. To achieve
this, define the absolute value of the log of the tolerance to be
$\gamma$ (25 for the example here) and solve $\sigma^{2}k^{2}/2+\alpha\left|k\right|=\gamma$.
This is a simple quadratic equation, but because one doesn't know
in advance how dominant the relative terms are, one should solve it
with a bit of care. The stable quadratic solution presented in Numerical
Recipes \citep{NumericalRecipes} can be adapted to be\begin{equation}
k_{max}=\frac{2\gamma}{\alpha+\sqrt{\alpha^{2}+2\gamma\sigma^{2}}}\label{eq:kmax}\end{equation}
This is simpler than the full solution in \citep{NumericalRecipes}
since the signs of both $\alpha$ and $\sigma$ are known in advance. 

Now, note that the periodic solution is really an infinite comb of
functions shaped like the desired one, added together. Since, beyond
a few $\sigma$ from the center, the function is close to Lorentzian,
one has really computed the desired function plus an infinite series
of offset simple Lorentzians:\begin{equation}
V(\alpha,\sigma;\, x)=V_{act}(\alpha,\sigma;\, x)+\frac{\alpha}{\pi}\sum_{n\neq0}\frac{1}{(x-n\Delta)^{2}+\alpha^{2}}\label{eq:vcorr1}\end{equation}
where $\Delta$ is the period of the function. However, the infinite
sum can be computed analytically. It is:\begin{equation}
\epsilon\equiv\frac{\alpha}{\pi}\sum_{n\neq0}\frac{1}{(x-n\Delta)^{2}+\alpha^{2}}=\frac{\sinh\frac{2\pi\alpha}{\Delta}}{\Delta\left(\cosh\frac{2\pi\alpha}{\Delta}-\cos\frac{2\pi x}{\Delta}\right)}-\left(\frac{\alpha}{\pi}\right)\frac{1}{x^{2}+\alpha^{2}}\label{eq:vcorreqn}\end{equation}

\begin{equation}
\frac{\partial\epsilon}{\partial\alpha}=\frac{\left(\alpha^{2}-x^{2}\right)}{\pi\left(\alpha^{2}+x^{2}\right)^{2}}+\frac{2\pi\cosh\left(\frac{2\alpha\pi}{\Delta}\right)}{\Delta^{2}\left(\cosh\frac{2\pi\alpha}{\Delta}-\cos\frac{2\pi x}{\Delta}\right)}-\frac{2\pi\sinh^{2}\left(\frac{2\alpha\pi}{\Delta}\right)}{\Delta^{2}\left(\cosh\frac{2\pi\alpha}{\Delta}-\cos\frac{2\pi x}{\Delta}\right)^{2}}\label{eq:dalphacorreqn}\end{equation}
Since the derivative with respect to $\sigma$ is very localized,
and falls to zero rapidly at the boundaries, no correction is needed
for it.

Although these equations look computationally intensive, they are
not so at all. Note that the $\cosh$ and $\sinh$ terms are of a
constant, and not evaluated at each point. Also, for the usual Fourier-space
case, $\Delta=2x_{max}$ so the $\cos$ term is just an evaluation
of the cosine from $-\pi$ to $\pi$ on the same grid the rest of
the function will be evaluated. If the Voigt function is to be evaluated
for many different $\alpha,\,\sigma$ pairs (as is the case in fitting
routines), but always on a grid with a fixed number of points, this
cosine only gets evaluated once, too, and can be cached for reuse.
Also, the correction and its derivative with respect to $\alpha$
share most of their terms in common, so this correction is really
a simple algebraic adjustment to the raw function table.

The correction term in eq. \ref{eq:vcorreqn} is an approximation
based on all the other nearby peaks being entirely Lorentzian, and
works well. However, it can be improved by a scaling argument, which
works significantly better. The non-Lorentzian nature of the correction
is due to the convolution of the Gaussian with the curvature of the
Lorentzian causing a slight widening even on the tails. Note that
convolution of a function with a Gaussian only affects even derivatives
(by symmetry), so the second derivative term is the lowest order this
could affect. Also note that this effect is getting bigger as one
approaches the next peak over (the edge of the boundary). Thus, one
can try a correction of multiplying the right hand side of eq. \ref{eq:vcorreqn}
by $1+ax^{2}$ where $a$ is to be determined. Much of the structure
of $a$ can be obtained by scaling, and it should be $a\propto\sigma^{2}/\Delta^{4}$.
Empirical testing has shown that a constant of 32 appears optimal,
so an improvement on eq. \ref{eq:vcorreqn} is:\begin{equation}
\epsilon=\left[\frac{\sinh\frac{2\pi\alpha}{\Delta}}{\Delta\left(\cosh\frac{2\pi\alpha}{\Delta}-\cos\frac{2\pi x}{\Delta}\right)}-\left(\frac{\alpha}{\pi}\right)\frac{1}{x^{2}+\alpha^{2}}\right]\left(1+\frac{32\sigma^{2}x^{2}}{\Delta^{4}}\right)\label{eq:vcorreqn-extra}\end{equation}
This improves the original correction by almost an order of magnitude
in the peak error at the bounds of the interval, and the RMS error
is reduced by about a factor of 5 for most test cases.

Note that this expression also provides an error estimate for the
calculation, which can be used to determine an appropriate value for
$\Delta$.  Assuming the error is of the same order as the final correction
term, which should be conservative, one can proceed to evaluate it
at the boundary of the interval, where $x=\Delta/2$, and evaluating
eq. \ref{eq:vcorreqn} assuming $\Delta\gg\alpha$, the error estimate
is then 

\begin{eqnarray}
\textrm{relative\, err} & \sim & \textrm{(small\, coefficient)}\times2\,\sigma^{2}/\Delta^{2}\label{eq:errorestimate}\end{eqnarray}
Note that, although this is $\propto1/\Delta^{2}$, this is the relative
error. The function itself is decreasing at the boundary as $1/\Delta^{2}$
so the absolute error scales with $1/\Delta^{4}$, as expected.

\section*{The road not taken}

There is another way one could consider carrying out this computation,
which looks elegant and easy from the outset, but actually is computationally
much more expensive. I will outline it here as a warning to others.

Instead of fixing the periodicity error by adding on the correction
of eq. \ref{eq:vcorreqn} or eq. \ref{eq:vcorreqn-extra}, one might
be tempted to fix the problem in advance, before the transforms are
carried out from $k$-space to real space. The obvious solution is
to try to compute the transform of the difference between the Voigt
function and a pure Lorentzian, and then add the pure Lorentzian back
in afterward, not as an infinite sum as in the correction equations,
but just as a single copy. One would compute\begin{equation}
\tilde{V}_{0}(\alpha,\sigma;\, k)=\left(\exp\left[-\sigma^{2}k^{2}/2\right]-1\right)\exp\left[-\alpha\left|k\right|\right]\label{eq:badxform}\end{equation}
and then transform this, and add back on the Lorentzian which was
subtracted. This turns out to be computationally very inefficient,
though. When computing the transforms, one wants a cleanly band-limited
function in $k$-space, with the power in the highest frequency channels
vanishing rapidly. In the case of eq. \ref{eq:convfft}, this is clearly
the case, since the $-\sigma^{2}k^{2}/2$ term makes the exponential
disappear relatively rapidly even for fairly modest values of $\sigma$
and $k$. In the case of eq. \ref{eq:badxform}, though, $\tilde{V}_{0}$
only vanishes as fast as the $\exp\left[-\alpha\left|k\right|\right]$
term, which falls off much more slowly. Thus, one has to carry out
the transforms to much higher values of $k$ to get convergence. This
turns out practically to be a huge penalty. Even in the case of $\sigma\approx k$,
it requires about a few times more terms, and in the case $\sigma\gg k$,
it is much worse, since the extremely rapid falloff of the Gaussian
in $k$-space allows one to sample only quite small values of $k$
to get very good performance.

\section*{Application}

The most probable way the author sees these gridded functions being
used is to load cubic spline interpolation tables to generate values
at points which may not lie on the grid. This way, one can compute
the Fourier transforms on grids of sizes convenient for FFT algorithms
(often, powers of two, but using. e.g. FFTW \citep{FFTW05}, many
other grid sizes can be conveniently transformed), and then use the
interpolator to fill in the values desired. Because of both the shape
of these functions and the wide dynamic range they typically encompass,
it is likely that it will be useful to interpolate the logarithm of
the function. Especially if $\sigma\gg\alpha$, so the center looks
Gaussian, log interpolation is extremely beneficial, since the logarithm
of the Gaussian part is just parabolic, and exactly interpolable by
a cubic spline interpolator. 

In fitting work, one more derivative is needed than the ones computed
in eqs. \ref{eq:ddalpha} and \ref{eq:ddsigma}, which is the derivative
with respect to the coordinate itself, which is needed to solve for
the center of a peak. Although this could easily be computed in Fourier
transform space by inverse transforming $i\, k\,\tilde{V}(k)$, it
also falls directly out of the cubic spline interpolation. The value
of a splined function at a point is computed (see the discussion of
splining in \citep{NumericalRecipes} for notation) as\begin{eqnarray}
h & = & x_{1}-x_{0}\nonumber \\
a & = & \frac{x_{1}-x}{h}\nonumber \\
b & = & 1-a\nonumber \\
y & = & a\, y_{0}+b\, y_{1}+\frac{h^{2}}{6}\left[(a^{3}-a)\, y_{0}''+(b^{3}-b)\, y_{1}''\right]\label{eq:yspline}\end{eqnarray}

then the first derivative is\begin{equation}
y'=\frac{y_{1}-y_{0}}{h}+\frac{h}{6}\left[(3\, b^{2}-1)\, y_{1}''-(3\, a^{2}-1)\, y_{0}''\right]\label{eq:ysplineprime}\end{equation}

This is the method preferred by the author of this work for this derivative.

\section*{Error Analysis}

\begin{figure}[p]
\noindent \begin{centering}\includegraphics[width=1\columnwidth]{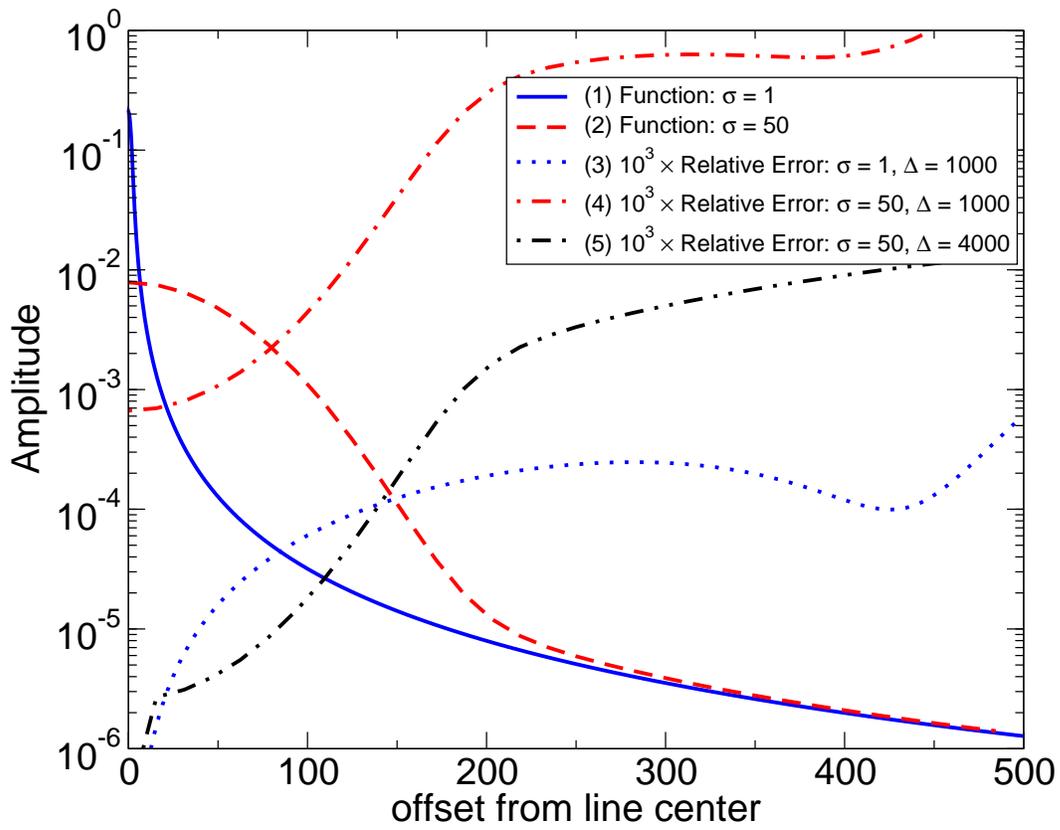} \par\end{centering}

\caption{\label{cap:ErrorPlot} Plot of Voigt Functions with $\alpha=1$ evaluated
by this method, and the relative errors (as computed by adaptive numerical
integration of the convolution at each point). }
\end{figure}

Figure \ref{cap:ErrorPlot} shows sample functions computed by this
method, and the relative errors associated with this computation.
These were computed using the extra error correction of eq. \ref{eq:vcorreqn-extra}.
The scaling of the errors is fairly easy to compute from the underlying
equations. In general, the errors scale as $\Delta^{-4}$ when $\sigma$
and $\alpha$ are held constant. For most practical applications,
it is likely that the need to compute the tails far enough from the
center that the entire spectrum is covered by the calculation results
in the tails being calculated sufficiently far out that the accuracy
is not a concern. In Figure \ref{cap:ErrorPlot}, the curve (5) shows
the result when $\Delta=80\sigma$ (tails computed to $40\sigma)$,
and even in this case the peak relative error is $10^{-4}$ (in a
part of the curve off the graph). Most practical cases are likely
to need the tails much farther out than this, resulting in the accuracy
automatically being better than this. As an example, though, of pushing
the computation to very narrow tails, the curve (4) shows the results
for the tails being computed to only $10\sigma$. Even in this case,
the relative error only exceeds $10^{-3}$ at the very edge of the
domain.

Note that these are computed with $\alpha=1$ and $\sigma$ being
varied. Since the shape of the Voigt function really only depends
on $\alpha/\sigma$ (related to the usual $y$ parameter), this completely
specifies the shape of the function except for an overall width scale.
Also, note that the cases shown are all for $\sigma\ge\alpha;$ the
case of small $\sigma$ is a limiting case, and the errors look much
as they do for $\sigma=\alpha$. 

\section*{Conclusion}

Because of relatively slow convergence, simple FFT-based convolution
has not fared well in the Voigt-function computation arena. Nonetheless,
this method has always had an advantage in simplicity and vectorizability.
Also, it is trivial to get the derivatives of the Voigt function with
respect to its width parameters from transform-based methods. This
makes this algorithm most useful for cases in which line strengths,
widths, and positions are all variable. The convergence enhancement
can also be easily extended to line shapes in which a non-Gaussian
function is convolved with a Lorentzian.

By combining the traditional transform-based method with a convergence-\-en\-hanc\-ing
operation, the result is a method which is fast, accurate, and extremely
easy to implement. It should find particular application for fitting
work carried out in many widely used scripting languages, in which
fast vector operations often make computation of tables of function
values an efficient process. As an example, on my 1 GHz laptop computer,
it takes 8 milliseconds to compute a 2048 point grid of the function
and its two derivatives using this method, in the Python language.
Even in compiled languages, though, this may be highly adaptable to
fast work on any operating system and machine which provides good
vector operation and FFT support. 

A more detailed speed comparison of this to other algorithms, in other
languages, with differing numbers of points, and differing accuracy
requirements, is difficult, since every one of these parameters affects
the speed of one algorithm relative to another. Compared to explicit,
point by point, computation of the Voigt function in a scripting language
(which was done, for Fig. \ref{cap:ErrorPlot}, by direct convolution)
it can be 2 orders of magnitude faster. Compared to the unenhanced,
transform-based algorithm of \citet{karp1978}, this provides much
more accuracy for the same speed, or many times better speed for the
same accuracy, where 'many' can often be a factor of 10. 

\newpage{}

\section*{References}

\renewcommand{\bibsection}{}

\bibliographystyle{unsrtnat}
\bibliography{/Volumes/Archive/marcus/BiblioDatabases/ebit}

\end{document}